\documentclass[conference]{IEEEtran}
\IEEEoverridecommandlockouts

\newif\ifsubmit
\submittrue

\newif\ifblinded
\blindedfalse

\usepackage{todonotes}
\presetkeys{todonotes}{inline}{}
\ifsubmit

\newcommand{\removed}[1]{}
\newcommand{\DV}[1]{}
\newcommand{\LV}[1]{}
\newcommand{\TODO}[1]{}
\newcommand{\CRITICAL}[1]{}
\else

\newcommand{\removed}[1]{}
\newcommand{\DV}[1]{\todo[inline,color=green]{DV: #1}}
\newcommand{\LV}[1]{\todo[inline,color=cyan]{LT: #1}}
\newcommand{\TODO}[1]{\todo[inline,color=orange]{TODO: #1}}
\newcommand{\CRITICAL}[1]{\todo[inline,color=red]{CRITICAL: #1}}
\fi

\usepackage{stfloats}
\usepackage{float}
\usepackage{caption}
\usepackage{tabularx} 
\usepackage{cite}
\usepackage{amsmath,amssymb,amsfonts}
\usepackage{algorithmic}
\usepackage{graphicx}
\usepackage{textcomp}
\usepackage{xcolor}
\usepackage{siunitx}
\usepackage[hidelinks]{hyperref}
\usepackage{cleveref}
\usepackage[normalem]{ulem}
\usepackage{pgfplots}
\pgfplotsset{compat=1.18}
\usepackage[table]{xcolor}
\usepackage[letterpaper, margin=0.75in]{geometry}
\usepackage{orcidlink}

\newcommand{\MHz}[1]{#1~\si{\mega\hertz}}

\newcommand{\GBs}[1]{#1~\si{\giga\byte\per\second}}

\begin{document}

\title{SUSpMV: A High Frequency Sparse Matrix Vector Multiplier on HBM Enabled FPGA written in SUS}


\ifblinded
\else
\author{\IEEEauthorblockN{Lennart Van Hirtum~\orcidlink{0000-0002-6982-8012
}}
\IEEEauthorblockA{\textit{PC\textsuperscript{2}, University Paderborn} \\
Paderborn, Germany \\
lennartv@mail.uni-paderborn.de} 
\and
\IEEEauthorblockN{David Volz~\orcidlink{0000-0002-1031-2695}}
\IEEEauthorblockA{\textit{ESA, TU Darmstadt} \\
Darmstadt, Germany \\
volz@esa.tu-darmstadt.de} 
\and
\IEEEauthorblockN{Andreas Koch~\orcidlink{0000-0002-1164-3082}}
\IEEEauthorblockA{\textit{ESA, TU Darmstadt} \\
Darmstadt, Germany \\
koch@esa.tu-darmstadt.de} 
\and
\IEEEauthorblockN{Christian Plessl~\orcidlink{0000-0001-5728-9982}}
\IEEEauthorblockA{\textit{PC\textsuperscript{2}, University Paderborn} \\
Paderborn, Germany \\
christian.plessl@uni-paderborn.de} 
}
\fi

\maketitle

\begin{abstract}
SUSpMV is a Sparse Matrix Vector multiplication (SpMV) accelerator written in the upcoming HDL \textit{SUS}.
By leveraging SUS's unique latency counting and inference mechanism, SUSpMV could be designed with very deep pipelines yet small design complexity overhead.
This enables an efficient implementation which employs all 32 HBM channels on the Alveo U280 FPGA at \MHz{400} for streaming matrix data into 32 Compute Units (CUs).
Input and output vectors are stored in DDR memory, which allows zero-overhead chaining of multiplications and increases overall system memory bandwidth by not sharing HBM bandwidth with the CUs.

A CU processes the SpMV in tiles of width 1024 and a dynamically chosen height, up to 32768.
Each CU is capable of accumulating the multiplications with up to 6 separate matrix entries per cycle, resulting in the combined theoretical peak computational throughput of 153.6 GFLOPs.
The matrix storage format is designed to exploit density variation within a given matrix by dynamically switching between one representation optimized for denser regions, and a second optimized for sparser regions, allowing jumps of up to 255 rows between each entry.

Evaluation demonstrates a 79\% geometric mean improvement over prior work on the same platform. We achieve a peak throughput of 144.9 GFLOPs, or 94\% of our theoretical computational throughput, compared to 98 GFLOPs reached by prior work.
\end{abstract}

\section{Introduction}

In a Sparse Matrix-Vector multiplication (SpMV), a sparse matrix is multiplied with a dense vector, resulting in a new dense vector.
It is a critical operation in a multitude of domains including linear algebraic solvers \cite{intro_solver1}, graph computation \cite{intro_solver2}, scientific computing \cite{intro_science1, intro_science2}, and machine learning \cite{intro_ml1, intro_ml2}.
With an arithmetic intensity of one multiply and one add per matrix element, the SpMV problem will be limited by the bandwidth of reading the matrix elements. Therefore implementations tend to focus on techniques to minimize the memory footprint of the matrix. To this end, most implementations across devices use standard sparse matrix formats such as Coordinate List (COO), Compressed Sparse Row (CSR) or Compressed Sparse Column (CSC) formats~\cite{hihispmv, serpens}.

Modern datacenter FPGA cards come equipped with High Bandwidth Memory (HBM)~\cite{fpga-spmv-efficiency, fpga-spmv-survey} interfaces.
An example is the Alveo U280, which offers 32 HBM channels with a combined total bandwidth of \GBs{460}. 
Newer devices such as the Alveo V80 extend this to \GBs{819}.

Due to this massive bandwidth, such FPGA cards prove to be a competitive alternative to modern CPUs and GPUs for memory-bound operations, even if they still fall far behind these platforms on raw number crunching due to their 10x lower operating frequency. That is, of course, if the available bandwidth is used well. 

A recent prior work that stands out in this regard is HiHiSpMV~\cite{hihispmv}, which independently improved on the earlier Serpens~\cite{serpens}.
HiHiSpMV was the first to utilize all 32 HBM pseudo-channels of the Alveo U280 at their full capacity, thus setting the bar for bandwidth utilization. 

Even though HiHiSpMV was able to exhaust the entire bandwidth of the U280 HBM, more efficient use of this bandwidth still seems possible. 
The free-form bitwise operations possible on FPGAs enable compression schemes that are difficult to decompress on other platforms.

A possible scheme for such bandwidth improvements is reordering elements and grouping them together to maximize utilization of bandwidth and compute resources \cite{XieGLZC025}.
However, when looking at sparse matrices, density often varies by multiple orders of magnitude between different regions even within the same matrix \cite{hispmv}.
Newer encodings can adapt to different densities, but so far required more resource-intensive accelerator designs \cite{cospmv}.
This is an area in which SUSpMV was designed to push the boundaries. A special encoding scheme was developed for SUSpMV which dynamically switches between a high-density variant and a low-density variant. Details on this are provided in \cref{sect:matrixFormat}. 


In many tile-based implementations the accumulation is usually limited to producing a single Y value per cycle. In such implementations, despite there being 8 or 16 floats in a block, if a given row only has one value, the remaining slots would go unused, as happens in \cite{b8c} and \cite{hihispmv}, or the accelerator itself would introduce backpressure like in \cite{fowers14}.
Both reduce the memory bandwidth. In designing SUSpMV we decided from the outset that accumulating to multiple rows within one cycle was a priority, and implemented it as such. Details are discussed in \cref{sect:partSum}. 

\removed{
\DV{this paragraph we have to rewrite}
Most recent works used HLS to develop the hardware accelerator, and while we agree that HLS can drastically ease the development of accelerators, it is our opinion that doing so both prevents certain interesting implementations of a problem, and leaves a lot of performance on the table. Be it in reduced FMax, or larger-than-needed resource consumption. Since we indeed have come up with an interesting design for the SpMV problem, we decided to implement it using the upcoming HDL SUS~\cite{sus-website}, since our design is uniquely suited to its unique Latency Counting and Latency Inference system. This SpMV design we dubbed "SUSpMV". }

Most recent works used HLS to develop the hardware accelerator, which can drastically ease implementation at the cost of reducing the developers' agency over the design's finer details. We implemented SUSpMV in the upcoming HDL SUS~\cite{sus-website}, since our design is well suited to its unique Latency Counting and Latency Inference system.

SUSpMV introduces a dynamic matrix storage format that is easy to decode on FPGA which dynamically adapts to high- and low-density ranges within the same matrix, without the need for filler values or other stall mechanisms (\cref{sect:matrixFormat}).

Furthermore, SUSpMV introduces a novel accumulator design, which can accumulate six values from up to six distinct rows, thus reducing the number of idle cycles for each adder (\cref{sect:partSum}).

Using these innovations, SUSpMV exceeds the effective bandwidth utilization of HiHiSpMV even though actual bandwidth is lower due to not reaching the full \MHz{450}.

To summarize, this work's contributions include:
\begin{itemize}
    \item A dynamic matrix storage format that can take advantage of matrices with non-uniform density distribution and is easy to decode on FPGA.
    \item A novel accumulator design capable of six Multiply-Accumulates per cycle per CU, where the input values may all come from the same row, all different rows, or any combination thereof.
    \item A comparison of throughput in GFLOPs with HiHiSpMV.
    \item An ablation study evaluating the contribution of our matrix format and accumulator design to overall performance.
\end{itemize}

\section{Background \& Related Work}

\removed{In this section we will provide context for the two major aspects in which SUSpMV differs from other implementations.
We first discuss the matrix data format, followed by how our accumulator differs from other implementation.
Finally, we give some background on how the structure of the FPGA has influenced SpMV designs.}

In contrast to dense matrices, sparse matrices contain a majority of zero entries.
The density, or portion of nonzeros, often ranges from $10^{-1}$ to $10^{-6}$.
Computing with, or even just storing these sparse matrices as dense matrices is often impractical.

\subsection{Data Format}
\label{sect:backgroundDataFormat}

Sparse matrices are usually stored in formats such as Coordinate list (COO)~\cite{hispmv}, where each nonzero entry is stored as a value, row, and column triplet, Compressed Sparse Row (CSR)~\cite{hihispmv}, which stores them as (value, column) tuples with row pointers marking the start of a new row, or Compressed Sparse Column (CSC), which does the same with axes transposed.

While simple, these formats waste memory bandwidth by missing out on compression opportunities.
R3COO~\cite{XieGLZC025} improves on the COO format, by encoding ten values in 512-bit and consolidating control signals and compressing bit widths for row and col indices.
COSMAT~\cite{cosmat} proposes the OCOO format, a combination of COO and CSR where row indices are compressed to single bit entries indicating that following entries are moved to the next row.

\removed{
HiSpMV~\cite{hispmv} uses a COO format combined with tiling. Each matrix entry requires 64-bit in total.
\cite{XieGLZC025} improves on this by introducing the R3COO format, which encodes ten values in 512-bit by consolidating control signals.
HiHiSpMV~\cite{hihispmv} builds upon the CSR format.
\DV{Finish these sentences}
}

Accelerator design depends heavily on the chosen data format as it influences predictability of the memory accesses for the X and Y vectors.
When using a CSR format, accesses to the X vector are random, while accesses to the Y vector are linear.
CSC format reverses this, while COO format requires random accesses to both.
Using random accesses to external memory negatively impacts the available bandwidth.
For this reason, the corresponding vectors are usually buffered in local SRAM resources.
The DRC format \cite{lll23} optimizes for random X vector accesses, by grouping matrix elements using the same X value together where possible, thus enabling smaller buffer sizes.

When matrix dimensions exceed the capacity of these SRAM buffers, the matrix must be split into smaller \textit{tiles} that fit the accelerator and are processed separately. A limiting factor on the size of these tiles is therefore the size of the available SRAM. 
Tiles are usually also the granularity at which parallelism is implemented, with different CUs processing separate tiles in parallel, and merging their results afterwards. 

A dynamic matrix storage format that can adapt to different densities within a matrix was implemented by CoSpMV~\cite{cospmv}.
Its SCDF format is separated into blocks for control signals, values, and coordinates, which reside in separate address segments.
Since control and coordinates reference other blocks, the accelerator design is resource-intensive and, as a result, could not make use of all the available bandwidth to external memory.

\subsection{Accumulation}
\label{sect:backgroundAccumulation}

The accumulator is the main computational component of any SpMV accelerator.
It receives a parallel input stream of the products of X vector and matrix values and adds them to the current value stored for the corresponding Y vector values.

The chosen data format impacts the accumulator design.
In a CSR format, the input stream will contain many products for the same row.
In a CSC format, all products must be added to different Y values of different rows.
When using the COO format, arriving entries may not belong to the same row or column.
If the accumulator can only handle one row at a time, this requires a stall mechanism or a routing network to distribute each row to a specific accumulator \cite{bldl23}.
If the matrix entries are not distributed evenly across rows and columns, this may cause parts of the accumulator to be idle.

The R3COO format from \cite{XieGLZC025} uses multiple adders per HBM channel, each of which handles its own rows.
The incoming matrix values are reordered such that each accumulator receives one value per cycle.
If one accumulator has fewer elements to process, the remaining slots are filled with zeroes.

HiHiSpMV~\cite{hihispmv} uses a hierarchical approach to accumulation within a row.
Hence, all values entering the accumulator in a given cycle belong to the same row.

Highly imbalanced sparse matrices may have rows with several times as many entries than average.
In HiSpMV~\cite{hihispmv}, a row can be split onto multiple CUs and their results accumulated afterwards for the resulting Y value.

\subsection{Platform and Implementation}
\label{sect:backgroundPlatform}

Recent work primarily targets the Xilinx Alveo U280 and similar devices \cite{leam}, \cite{serpens}, \cite{hispmv}, \cite{hihispmv}.

The Alveo U280 offers three super logic regions (SLRs) also shown in \cref{fig:floorplan}.
The bottom SLR0 offers 32 HBM pseudo-channels, one DDR controller and one PCIe controller.
This complicates larger designs using all 32 HBM ports, because the PCIe controller uses much of the logic resources on the left side of SLR0.
Each HBM port can be accessed in a 256-bit or 512-bit configuration, saturating the bandwidth at 450 and \MHz{225} respectively \cite{fpga-hbm}.
With the exception of HiHiSpMV, no other design saturates HBM bandwidth.
The mid SLR1 offers a second DDR controller, which so far has not been used by SpMV accelerators.
Finally, the top SLR2 offers no peripherals relevant to SpMV.
Since most free logic resources are on SLR1 and SLR2, larger designs must be designed with costly SLR crossings in mind and contend with more aggressive floorplanning.

Other SpMV implementations commonly use HLS to develop the kernel \cite{spica} \cite{serpens} \cite{hispmv} \cite{hihispmv} \cite{fpga-spmv-efficiency}.
While HLS can reduce development cost, we believe SpMV to be a critical application that can benefit from the designer having more control over the generated hardware.

\removed{We use the upcoming HDL "SUS"~\cite{sus-website} for this project.
Latency counting is a unique feature of SUS and allows the designer to easily adjust pipelining in the design. The user can insert pipeline registers where needed, and the Latency Counting system will automatically insert registers on parallel paths to keep them in sync. An instance of where this is very relevant is in the design of the partial sums module (\cref{sect:partSum}). The fp32\_add sub-modules used to add the terms together each have a latency of 11 cycles. So after the terms have become available from the multipliers, the first adder can start immediately, the second adder has to wait 11 cycles, the third 22, etc. Latency Counting automatically adds the necessary registers to introduce these delays so we as designers don't need to worry about it.

The "reverse" of Latency Counting is Latency Inference, where the compiler uses the pipelining information it sees in the logic around a module, and from this determine one or more latency-sensitive parameters on the module. In our case, Latency Inference figures out that the ALMOST\_FULL\_THRESHOLD if the CU Matrix Data input FIFOs on SLR0, SLR1, and SLR2 should be 204, 212, and 228 respectively. (192 cycles are introduced by the axi3\_burst\_reader to account for in-flight AXI read bursts, and 12, 20, and 28 pipeline stages introduced to cross the SLRs.)\footnote{More on this \url{https://sus-lang.org/docs/docs/latency_counting/latency_inference.html}}

Signal names in SUS map 1:1 to their counterparts after compiling the design to SystemVerilog.
This enables the designer to do low-level optimizations, such as mapping to Xilinx primitives, directly in SUS.}

We use the upcoming HDL "SUS"~\cite{sus-website} for this work.
Latency counting is a unique feature of SUS and allows the designer to easily adjust pipelining in the design. The user can insert pipeline registers where needed, and the Latency Counting system will automatically insert registers on parallel paths to keep them in sync. This is very relevant in the design of the accumulator, for instance (\cref{sect:partSum}). The 32-bit floating point adders to accumulate the six multiplied terms have a latency of 11 cycles each. As a result, the first adder can start immediately, the second adder must wait 11 cycles, the third 22, etc. Latency Counting automatically adds the necessary registers to introduce these delays.

The "reverse" of Latency Counting is Latency Inference\footnote{More on this \url{https://sus-lang.org/docs/docs/latency_counting/latency_inference.html}}, where the compiler uses the pipelining information it sees in the logic around a module, and from this determine one or more latency-sensitive parameters on the module.
Using this feature, modules can dynamically adapt to the latencies of surrounding modules.

\section{Implementation}

For the SpMV problem, the key bottleneck remains memory bandwidth. Any bits that we don't put to use in the encoding directly count against the performance we could reach. 
There are two ways in which we may waste bits:
The first is inefficient compression of data, such as using 32 bits for X indices. 
The second is when the accelerator architecture requires the insertion of zeros, which is the case if a block can only accumulate to one row. All remaining slots in such a block would have to be set to zero, which can add up to a significant amount of unused bandwidth. 
Our solution to the first problem is the 5xFloat/6xFloat format; an encoding scheme that either packs 6 floats, with 6 10-bit X indices into a single block for dense regions, or packs 5 floats with 5 10-bit X indices and 5 8-bit delta-Y values for sparse regions (\cref{sect:matrixFormat}).

Our solution to the second problem comes in the form of allowing each 5xFloat/6xFloat block to have "go to the next row" markers at any position, allowing a single block to contribute partial sums to up to 6 rows at the same time. To support this, the Partial Sums system (\cref{sect:partSum}) has been designed to support arbitrary cuts between any pair of elements, and likewise the Y Collector was designed to ingest up to 6 partial sums per cycle (\cref{sect:y_collector}).

\subsection{The Matrix Format}
\label{sect:matrixFormat}

\begin{figure}
    \centering
    \includegraphics[width=1\linewidth]{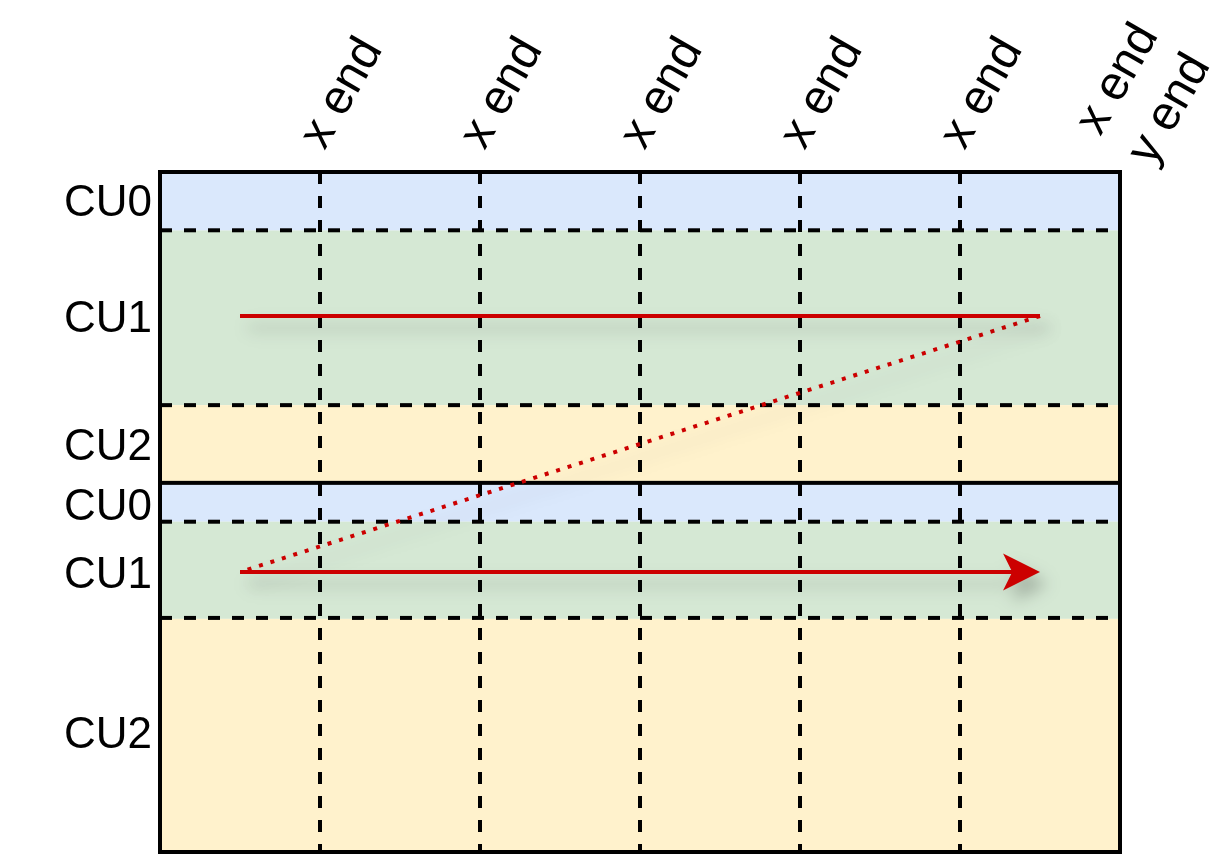}
    \caption{Tile Iteration Structure. For the sake of this example the Y axis was split into 6 rows of tiles, with a roughly equal number of nonzeros each. If we have three CUs available, we will set Y repeats to 2, such that each CU handles two sets of 6 tiles, for a total of 6 tiles in the Y direction. The X vector in DDR memory will be read exactly twice, once for the first three rows of tiles, and once for the second three rows of tiles.}
    \label{fig:tilingIteration}
\end{figure}

The matrix is split into tiles of width 1024 X values, and up to 32768 Y values high. The width of a tile is always fixed at 1024 X values, whereas the height of the tile is allowed to vary up to a maximum of 32768. (An illustration of this is provided in \cref{fig:tilingIteration}) By varying the height of a row of tiles, we can distribute the computational effort as equally as possible. (For more details, see \cref{sect:computeUnitLoadBalancing}) All tiles in a row are handled by the same CU, and then these rows of tiles are then distributed among the CUs. If splitting the matrix height over all CUs would cause one CU to be responsible for more than 32768 rows, we have to perform multiple passes over the matrix. We call such a full pass a "Y repeat", and it corresponds to one full read of the X vector from DDR memory, which is broadcast to all CUs. 

\begin{figure}
    \centering
    \includegraphics[width=1\linewidth]{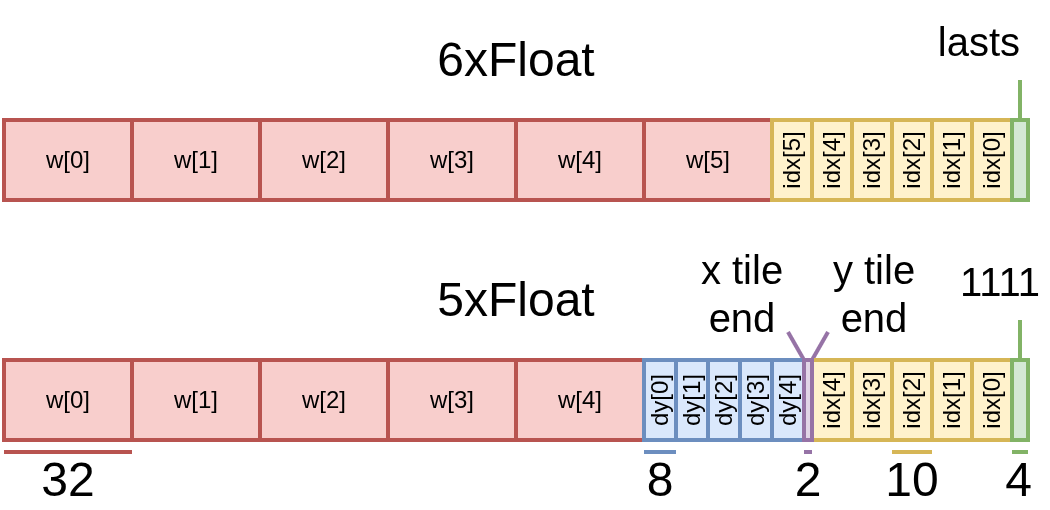}
    \caption{256-bit 5xFloat and 6xFloat compressed matrix data formats. The 6xFloat format is optimized for more dense sections of the matrix, by having space for 6 weights and 6 10-bit indices. The 5xFloat format (indicated by the 4 "lasts" bits being set) exchanges one weight in favor of storing delta-Y values to support greater sparsity. Likewise only the 5xFloat variant can encode the end of a tile. A usage example of these blocks is shown in \cref{fig:5xFloat6xFloatTile}}
    \label{fig:matrixFormat}
\end{figure}

Each tile is encoded as a sequence of 256-bit 5xFloat or 6xFloat blocks, which are streamed into each CU at one block per cycle.
The internal structure of the 5xFloat and 6xFloat formats are shown in \cref{fig:matrixFormat}. 
Each block encodes value and position of the next matrix entries within the current tile.
The X position within the tile is encoded with a 10 bit integer, while the Y position is relative to the previous entry.
For 5xFloat blocks, each entry can advance the Y position by 255, while 6xFloat is optimized for dense regions by trading those bits to store one more value and X position.
Entries in 6xFloat blocks can still cover multiple rows using either the last mask encoded in the last 4 bits, or when an X position is followed by a smaller X position, which also indicates a Y increment.
Together, those cannot represent all possible combinations, but exceptions are rare.
Throughout all matrices used in our evaluation, usage of a 6xFloat over a 5xFloat was prohibited in 0.17\% (geomean) of cases and 3.4\% (worst case) for this reason.
\removed{However, we note that regression to 5xFloat always happens for consecutive elements on a purely diagonal matrix.}

\begin{figure}
    \centering
    \includegraphics[width=1\linewidth]{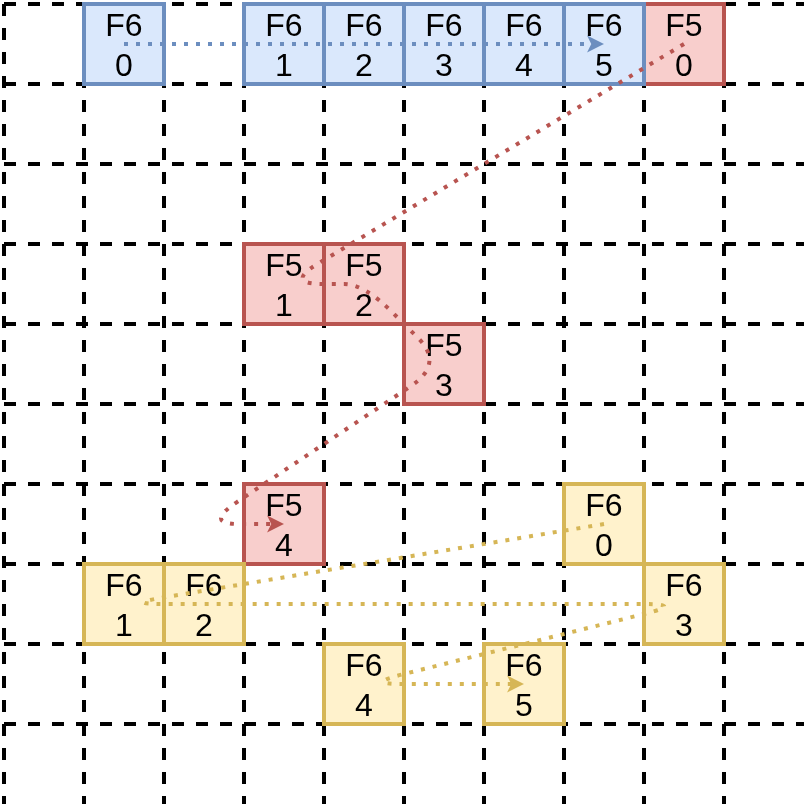}
    \caption{An example of the start of a tile. The 16 sparse values in this tile are shown as they would be stored in three HBM storage blocks. The first block is a 6xFloat tile covering elements in row 0, followed by a 5xFloat tile contributing the sparse values for row 0, 3, 4, and 5, finally followed by another 6xFloat block spanning rows 5 through 7.}
    \label{fig:5xFloat6xFloatTile}
\end{figure}

\removed{
\subsubsection{Extra Constraints on the Matrix Format}
\label{sect:extraMatrixConstraints}
}

\subsection{A single Compute Unit}
\label{sect:singleComputeUnit}
Observe on \cref{fig:computeUnit} the structure of a single CU. The data flows from the bottom upwards, with encoded 5xFloat/6xFloat values flowing in from the HBM from the bottom, flowing through the FIFO for backpressure, until being decoded in the Decoder into 6 weights, 6 X indices, 6 delta-Y values, and flags denoting the end of an X tile, or the end of a whole row of tiles. The vector values for the current tile are looked up in the BRAMs, before being multiplied with the weights. These multiplied terms then flow up through the Partial Sum computer, and Accumulator, at which point we are left with 0-6 valid partial row sums. These partial row sums are then routed to any of 16 Y Collector banks for cross-tile accumulation. Finally, when all tiles in a row have been processed, the total sums for each row are streamed out of the Y Collector banks, and the banks are reset to 0.0 ready for the next Y repeat. 

\subsubsection{Reading from HBM and Block Decoding}
\label{sect:blockDecoding}
\removed{
Each CU is connected to a single HBM bank for retrieving the Matrix Weights. The AXI4 readers and writers used are from \cite{sus-xrt}, which provides a cheap memory-less AXI4 burst reader. The 256-bit weights are pipelined across the SLRs and arrive at the FIFO at the bottom of the design. (\cref{fig:computeUnit}) This FIFO is used to buffer the data stream coming from the HBM, and through the free-flowing pipeline, since these cannot be stopped abruptly. The 256-bit encoded 5xFloat or 6xFloat variants are unpacked by the decoder into 6 X indices, 6 weights (with a final 0.0 weight inserted for 5xFloat blocks), 6 delta-Y values produced (with 6xFloat' last signals converted to $dy[i] = 1$, and finally X tile end, Y tile end signals. The flow control logic for the FIFO keeps track of how many X buffers are in use, how many Y buffers, and controls the flow accordingly. 
}
Each CU is connected to a single HBM bank for retrieving the matrix storage data. The 256-bit blocks are pipelined across the SLRs and arrive at the FIFO depicted at the bottom of \cref{fig:computeUnit}.
This First-Word-Fall-Through FIFO buffers the data stream coming from the HBM, because our bursting reader needs several cycles to empty its in-flight data when interrupted.
The encoded 5xFloat or 6xFloat variants are unpacked by the decoder into six X indices, six weights (with a final 0.0 weight inserted for 5xFloat blocks), six delta-Y values (with 6xFloat' last signals converted to $dy[i] = 1$), and finally tile end control signals.
The flow control logic keeps track of how many X buffers and Y buffers are in use and controls the CU accordingly.

\subsubsection{X vector memory}
\label{sect:xVectorMemory}
Since accesses to the X vector are random, we use SRAM buffers to improve effective memory bandwidth.
The X vector memory stores the 1024 fp32 X values associated with the current tile.
They are read once, and reused for all entries in this tile. 

It is implemented as an 8-way double buffer, which can be read from and written to asynchronously.
This means that while the accelerator is using the X values of one tile, the next 7 tiles worth of X values can already be read in.
This asynchronous loading of X tile data was chosen because all CUs share the X vector reader.
By having capacity for 8 tiles, we allow some desynchronization of the CUs to occur.
However, for very unbalanced matrices, a fast CU may have to wait on a slower CU.

Because a 6xFloat block provides 6 X indices, and 6 weights, we must replicate this memory 6 times to enable maximum throughput. 
\removed{As shown by the blue memory blocks in the bottom of \cref{fig:computeUnit} each of the 6 memory blocks is realized as 8 BRAM tiles on the U280, for a total BRAM consumption of 48 BRAM36K.
The use of 8 BRAMs per memory block not only grants us additional capacity, but also a write bitwidth of 512-bit, allowing us to write 16 X values per cycle, alleviating this potential bottleneck for very sparse matrices.}
Each of the X memory banks is 512-bit wide in order to store incoming X values at the full bandwidth of the DDR interface.

After reading the associated X values, we multiply them by the 6 matrix weights to get the 6 terms that are passed on to the accumulator.

\subsubsection{Partial Sums \& Accumulation}
\label{sect:partSum}

The partial sum computation is shown as the chain of adders in the middle of \cref{fig:computeUnit}.
It consists of two main components: computing the partial sums for any section of the 6 terms, and the cross-cycle accumulation. 

\begin{figure}
    \centering
    \includegraphics[width=1\linewidth]{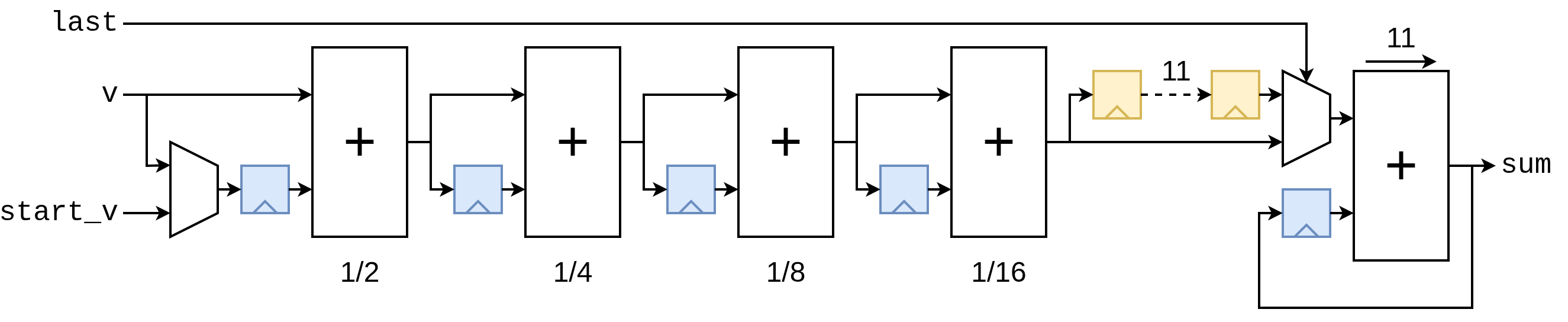}
    \caption{Cascading Accumulator}
    \label{fig:accumulator}
\end{figure}

For each block, the decoder will decode 6 delta-Y values.
These describe how many rows the accumulator must jump to the next nonzero.
A $dy[i] = 0$ value implies that the next element is part of the same row, and thus they should be accumulated together.
If $dy[i] \neq 0$, the accumulation should end at element $i$, and a new accumulation starting from 0.0 should begin.
This is implemented with the small multiplexers in \cref{fig:computeUnit}.
For row sums that extend beyond multiple blocks, a Cascading Accumulator (\cref{fig:accumulator}) is used.
It works by adding up adjacent pairs in stages, which slows down the data stream with each stage until the time between values is large enough to cover the 11 cycle latency for an fp32 addition.
\removed{(some care has to be taken to delay the "last" value, as it can arrive at the final adder before the previous sum of 16 values has finished its roundtrip).}
The accumulated sums, along with any partial sums that didn't need to pass through the Cascading Accumulator are passed up to the Y Collector, to be routed to their respective Y sum buckets. 

\subsubsection{Y Collector}
\label{sect:y_collector}

\begin{figure}
    \centering
    \includegraphics[width=1.0\linewidth]{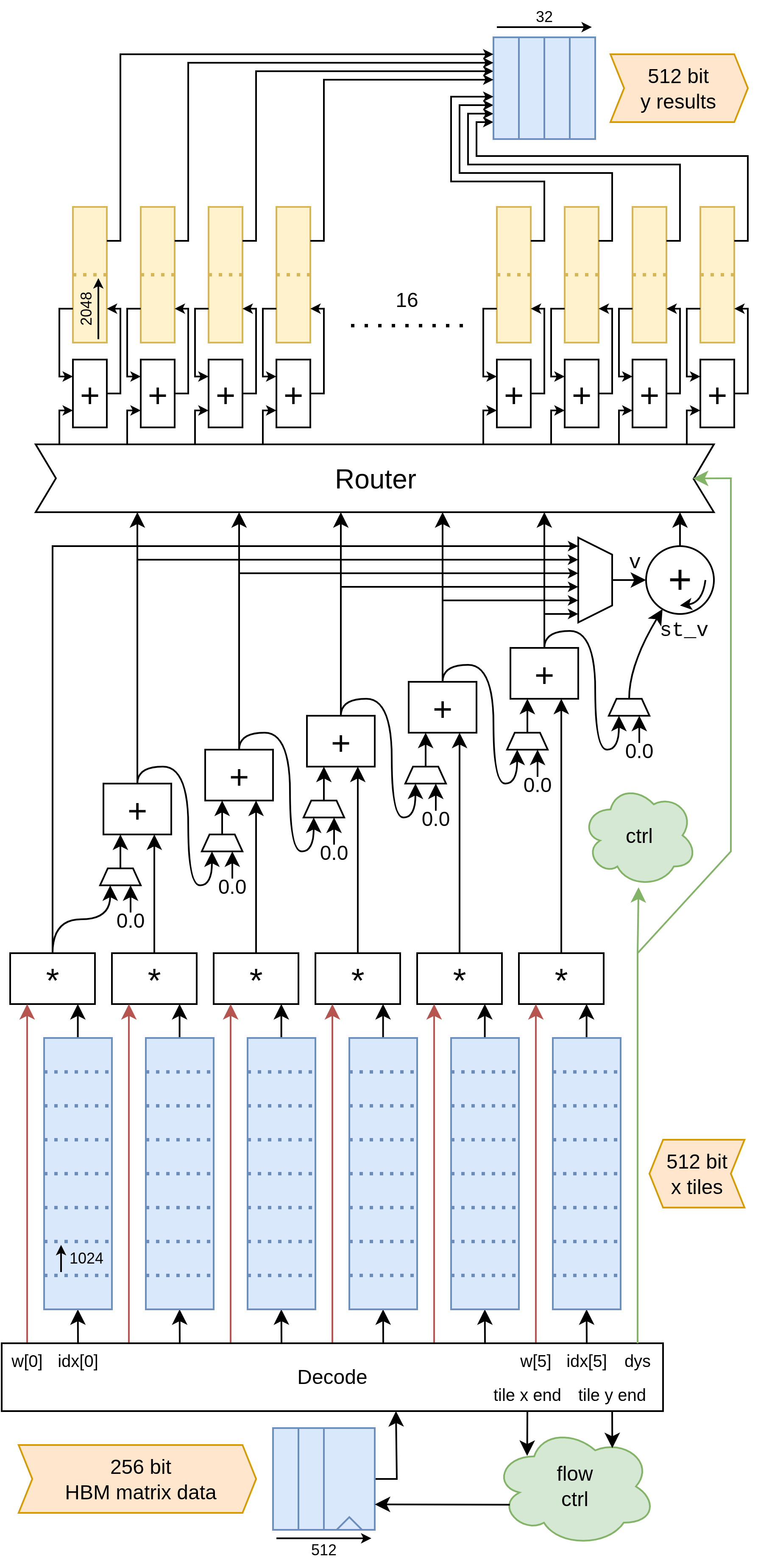}
    \caption{A single SUSpMV CU. Compressed matrix weights flow into the FIFO on the bottom, before being passed to the 5xFloat/6xFloat Decoder. The X values corresponding to the matrix weights are looked up and multiplied by the weights, after which partial sums are computed. Finished partial sums are routed to one of 16 Y value banks, and added onto the totals.}
    \label{fig:computeUnit}
\end{figure}

\removed{
The Y collector is responsible for carrying over the row accumulated values from one X tile to the next.
The Y vector results accumulated so far are each stored in one of 16 banks, implemented as one URAM each and shown as yellow memory blocks near the top of \cref{fig:computeUnit}.
From the accumulator of the pipeline we receive anywhere from 0 to 6 valid partial sums. From the Matrix Block Decoder we receive the delta\_y values, which we use to compute the rows that these partial sums belong to. We use the lower 4 bits of the resulting row index for each partial sum to choose the bank, and the remaining 11 bits index into one half of the URAM double buffer (2048 elements). 

This banking does mean we rarely have bank conflicts, which have to be resolved statically in the conversion to the HBM matrix storage format. It has to be ensured that no two partial sums within a single block share a Y index $mod 16$. To resolve this, the block must be split into two separate blocks.\removed{(More on this in \cref{sect:extraMatrixConstraints}.)}

Like the X vector BRAMs, the Y vector URAMs are structured as a 2-way double buffer. When all additions on the front buffer have been completed, the front buffer is released, which immediately allows the CU to start work on the next Y repeat. In the meantime, the back buffer can be slowly emptied. Because the URAM only has one read/write port, reads/writes from the main pipeline take priority, as to never backpressure the main pipeline, with the outputting to the FIFO and clearing of the memory contents using the unused cycles. 

To then recombine the 16 separate Y result streams and create batches of 16 floats that can be ingested by the DDR writer, a small "Joining" FIFO of depth 32 is used. 
}

The Y collector is responsible for carrying over the row accumulated Y values from one X tile to the next.
In a given cycle, it receives up to 6 partial sums together with their row indices.
The partial Y results accumulated so far are stored across 16 Y banks shown as yellow memory blocks near the top of \cref{fig:computeUnit}.
Banks are assigned by the lower 4 bits of the row index.
Any arising bank conflicts, e.g. when writing to row 0 and 16, must be resolved statically by moving the conflicting row to the next data block.

Once the CU finishes the last tile of a row, the accumulated Y results must be written to external memory.
This poses a problem, since SRAM resources on FPGA only offer two ports, both of which are occupied with accumulation.
We solve this by splitting each Y bank into a double buffer.
While one half is used for accumulation, the other half uses the idle cycles on that particular bank to stream out its data and reset its contents to zero.
The 16 streams are recombined and forwarded to external memory.

\subsection{Integrating 32 SpMV Units}
\label{sect:integrating32Units}

\removed{
In \cref{sect:singleComputeUnit} we discussed the structure of a single SpMV unit. This unit still has three interfaces that need to be connected:
\begin{itemize}
    \item The 256-bit encoded Matrix Data input
    \item The 512-bit wide X vector data input
    \item The 512-bit wide Y result data output
\end{itemize}

For the matrix data, each CU gets connected to its own HBM bank. The U280 has 32 HBM banks, and hence, the whole accelerator consists of 32 CUs. 

For the X vector and Y vector, our goal was to have them placed in a single memory, in part to have a simpler host code implementation, and in part not to share the HBM bandwidth with the matrix data. DDR1 was chosen for this purpose.

The X vector reader module is responsible for reading tiles of the X vector, and broadcasting them to all 32 CUs. Internally, it keeps track of how many tiles are in use in each CU. As discussed in \cref{sect:xVectorMemory} each CU can hold up to 8 tiles worth of X values in its local memory. When it has processed the last element of a tile, it emits a x\_tile\_release pulse to inform the X vector reader that a tile has freed up in this unit. 

Finally, the Y results are gathered via a shared bus back to the Y result writer. Control of this shared bus is managed using a token system. Initially, the Y result writer gives the bus token to CU 0. When it is ready, it starts writing its result values to the shared bus, and by asserting the "last\_y" flag on the last transfer it again relinquishes control of the shared bus. Once it sees this, the Y result writer grants the token to the next CU, in a round-robin fashion. 
}

To integrate multiple CUs into a single unified accelerator, their three data interfaces must be connected appropriately.
(i) The 256-bit encoded matrix data input uses a 1:1 connection to a single HBM pseudo-channel.
Because the Alveo U280 provides 32 such channels, our accelerator comprises 32 CUs.
(ii) The 512-bit wide X vector data input requires a 1:n broadcast.
The X reader streams this data from DDR memory to each CU.
Since CUs may progress at different rates, the X reader keeps track of their current state and stalls the input stream to prevent X values in the CUs' local X vector buffer from being overwritten too early.
After the entire X vector has been streamed in, the process repeats for every Y repeat.
(iii) The 512-bit wide Y result vector output requires a n:1 switch.
The Y writer joins all CU output streams with packet-based round-robin arbitration to ensure Y values are written in the correct order to DDR memory.

Finally, we use TaPaSCo~\cite{tapasco} to provide the PCIe host, DMA engine for transferring matrix and vector data, and control our accelerator via a memory-mapped register interface.
While TaPaSCo provides optional generic HBM access~\cite{tapasco-hbm}, we use a custom plugin to configure the HBM and its connections for our specific use-case.
It enables the HBM crossbar and connects TaPaSCo's DMA engine to one channel for uploading the matrix data.

\subsubsection{Floorplanning}

\begin{table}[h]
    \centering
    \begin{tabular}{c|c|c|c}
    Resource & One CU & SLR Left Half & SLR Right Half \\
    \hline
    BRAM & 48 & 384 & 288 \\
    URAM & 16 & 128 & 192 \\
    DSP & 28 & 1440 & 1440 \\
    LUT & $\sim$15000 & 220800 & 218880 \\
    FF & $\sim$16000 & 441600 & 437760 \\
    \end{tabular}
    \caption{Resource consumption of a single CU, as well as the total resources available on the left half (resp. right half) of each SLR. Note the strong imbalance in BRAM (RAMB36) and URAM resources available on the left half and the right half, with the left half being light on URAMs, and the right half being light on BRAMs.}
    \label{tab:computeUnitResourceUse}
\end{table}

\begin{figure}
    \centering
    \includegraphics[width=0.8\linewidth]{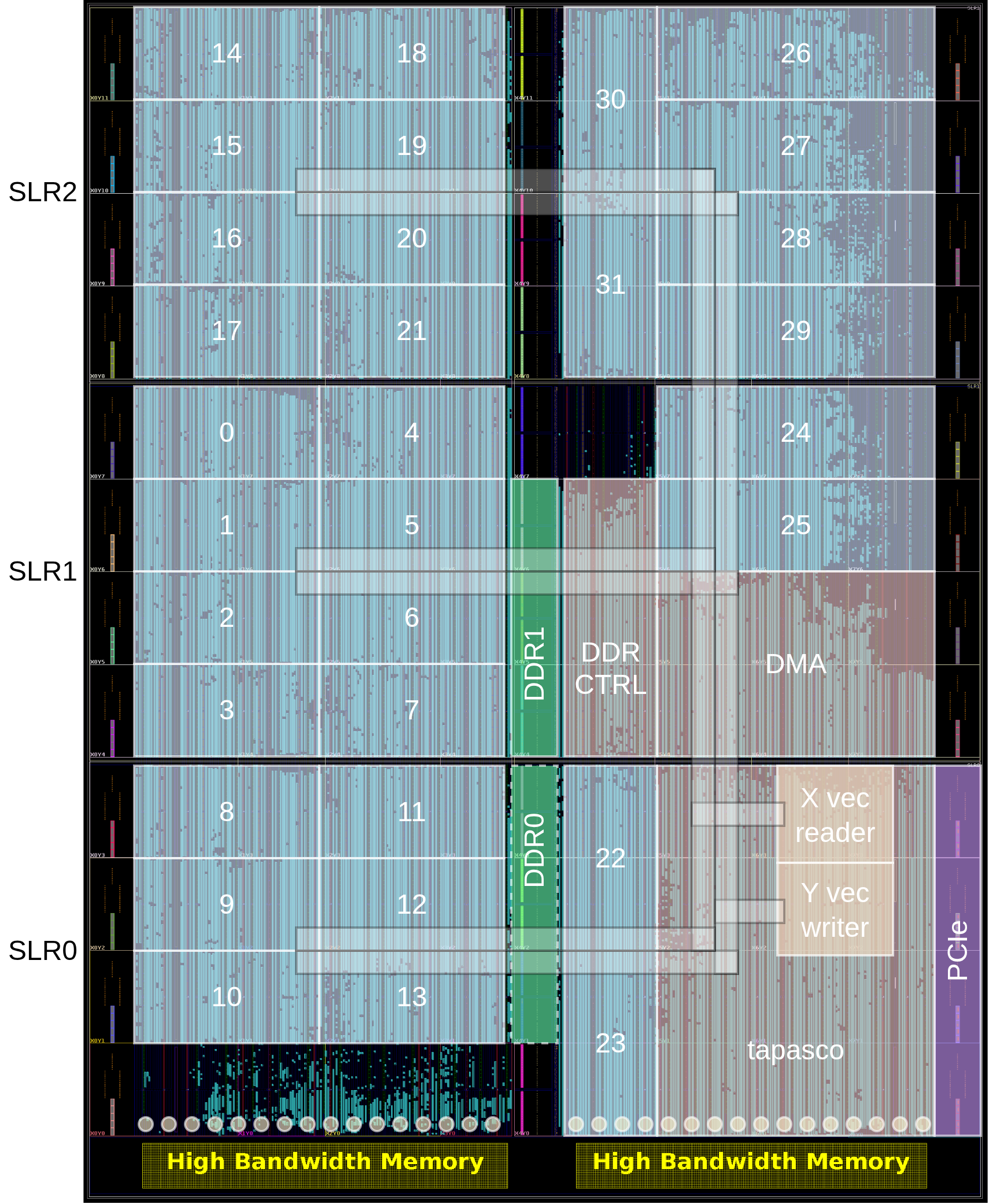}
    \caption{Floorplanning of the accelerator. Shown in yellow are the external interfaces that were used. In red the hardware for TaPaSCo. Green shows the shared modules in the accelerator: The AXI control slave for starting \& stopping of the kernel, and the X vector reader, and the Y vector writer which connect to DDR1. In blue, the distribution of the 32 CUs.}
    \label{fig:floorplan}
\end{figure}

\removed{
\Cref{tab:computeUnitResourceUse} shows the resource consumption of a single CU, as well as the resource availability on the left and right half of each SLR. Note that this means we can fit exactly 8 units on the left side of an SLR, and exactly 6 units on the right side of an SLR. When we do this, we reach 100\% utilization of BRAMs in that region of the FPGA, and 100\% utilization of URAMs on the left side. To aid Vivado in the placement - since otherwise it fails to find a decent placement of the BRAMs, we explicitly create a pblock for each CU with exactly the required amount of BRAMs and URAMs. On the left side, that leads to 8 nicely tiled units, whereas on the right side, to reach all needed resources 4 units had to stretch over 3 clock regions, with the other 2 units being spread out vertically. 

Another important consideration during floorplanning was long-range wires. Namely the wires for broadcasting the X vector from the X vector reader, the Y result channel going to the Y vector writer, as well as the connections between each CU and its corresponding HBM reader. The shared X vector bus and Y result bus are routed to the different half SLRs at which point fan out to the CUs. (See \cref{fig:floorplan}). Every one of these long buses is a free-flowing pipeline, of 8-24 pipeline stages depending on the number of SLRs visited. 
}

\Cref{tab:computeUnitResourceUse} shows the resource requirements of a single CU and available resources on the left and right half of each SLR.
Due to BRAM and URAM constraints, the left side of each SLR can fit exactly eight CUs (100\% BRAM and 100\% URAM usage), and the right side exactly six (100\% BRAM usage and 50\% URAM usage).

With such high utilization, Vivado struggles to achieve a reasonable design frequency.
Hence, we use pblocks to assign each CU to a specific region, containing exactly the required amount of BRAM and URAM resources as shown by the blue boxes in \cref{fig:floorplan}.

Another important consideration during floorplanning were long-range wires, namely, the wires for broadcasting the X vector and joining the Y vector (both shown in white), as well as the connections between each CU and its corresponding HBM port (not shown).
The shared X vector bus and Y result bus are routed explicitly to the different SLR halves, then fan out to local CUs using normal place and route. 
Each of these long buses is a free-flowing pipeline with 8-24 pipeline stages depending on the number of SLRs visited.

\subsection{Compute Unit Load Balancing \& Shuffling}
\label{sect:computeUnitLoadBalancing}

To ensure the computational effort is distributed equally across CUs, we employ two techniques.
First, we dynamically vary the height of the tiles such that each CU processes roughly the same number of matrix elements, as shown in \cref{fig:tilingIteration}.
If tile height exceeds 32k or the distribution is too imbalanced, we add additional Y repeats.
While this compensates for the general imbalance between rows, large-scale imbalances between the columns (such as a large void in one quadrant of the matrix), could cause some CUs to wait on other units blocking the loading of new x values.
To alleviate this second imbalance, we optionally employ random shuffling of rows and columns, as is also done by other works such as HiHiSpMV~\cite{hihispmv}.

\removed{
To account for density imbalances in the matrix, we employ two techniques for balancing the computational effort between the 32 CUs.
The first technique is dynamically varying the height of the tiles such that each CU processes roughly the same number of matrix elements, as shown in \cref{fig:tilingIteration}. If after such a partitioning one of the tile heights exceeds 32k, we increase the number of Y repeats and partition again. While this compensates for the general imbalance between rows, large-scale imbalances between the columns (such as a large void in one quadrant of the matrix), could cause some units to have to wait due to other units blocking the X vector reader. To alleviate this second imbalance, we optionally employ random shuffling of the matrix entries, in both the X and Y direction. Now, because shuffling changes the interpretation of the X and Y vectors, we treat with/without shuffling as separate methods in our evaluation in \Cref{tab:evalThroughputComparison}. Similar Shuffling was also performed in \cite{hihispmv}. 
}

\section{Evaluation}

We evaluate SUSpMV by comparing it to HiHiSpMV using their published results.
Afterwards, we present an ablation study to show the effectiveness of our dynamic matrix storage format and multi-row accumulator.

\subsection{Setup}

We integrated our SUSpMV accelerator into a complete design on the Alveo U280 using TaPaSCo~\cite{tapasco} with our custom plugin, then synthesized with Vivado 2023.2.

The host application uses the TaPaSCo C++ API for data transfer and accelerator configuration.

As with HiHiSpMV and other works, we only measure the kernel runtime and ignore pre-processing and data transfer.
We use a cycle counter for accurate measurement.
SpMV is usually employed in iterative algorithms, such that constant factors contribute little to overall runtime.

We measure time and matrix memory footprint, to report 32-bit float operations per second (GFLOPs) corresponding to $2*NNZ/time$, and effective bandwidth usage (GB/s) measured in nonzero matrix entries streamed by total time.

\removed{As we distribute the workload across 32 CUs, we employ random shuffling of the rows and cols to improve the distribution of nonzeros.
For this, we randomize the layout ten times, then report the best result.}

We use the same 59 matrices from SuiteSparse~\cite{suitesparse} as HiHiSpMV for each comparison.
\removed{We point out that HiHiSpMV reports higher density for \textit{X},  \textit{X},  \textit{X} matrices.}

\subsection{Throughput}

\begin{table*}[t]
\centering
\resizebox{\textwidth}{!}{%
\begin{tabular}{|r l r r r | r l r r r | r l r r r | r l r r r|}
\hline
ID & Matrix Name & HiHi & NoSh & Sh &
ID & Matrix Name & HiHi & NoSh & Sh &
ID & Matrix Name & HiHi & NoSh & Sh &
ID & Matrix Name & HiHi & NoSh & Sh \\
\hline

1  & human gene1  & 93.69 & 144.1 & \cellcolor{green!25}144.9 &
16 & TOPF1047     & 77.40 & 97.1  & \cellcolor{green!25}116.7 &
31 & olafu        & 44.19 & 62.0  & \cellcolor{green!25}91.7 &
46 & bcsstk24     & 30.42 & 36.5  & \cellcolor{green!25}36.8 \\

2  & human gene2  & 98.12 & 133.3 & \cellcolor{green!25}143.1 &
17 & appu         & 47.44 & 112.3 & \cellcolor{green!25}112.5 &
32 & nemeth20     & 60.78 & 66.9  & \cellcolor{green!25}94.6 &
47 & ted\_B       & 13.79 & 28.2  & \cellcolor{green!25}28.5 \\

3  & nd6k         & 80.13 & 101.5 & \cellcolor{green!25}137.1 &
18 & crystk03     & 33.00 & 46.5  & \cellcolor{green!25}103.6 &
33 & cbuckle      & 37.89 & 58.9  & \cellcolor{green!25}77.5 &
48 & bodyy6       & 4.64  & 18.5  & \cellcolor{green!25}20.2 \\

4  & TRS b2052    & 51.12 & 79.8  & \cellcolor{green!25}131.6 &
19 & nemeth26     & 72.75 & 68.8  & \cellcolor{green!25}108.9 &
34 & Trec13       & 58.97 & \cellcolor{green!25}87.7 & 87.5 &
49 & bodyy4       & 5.04  & 18.1  & \cellcolor{green!25}19.1 \\

5  & thread       & 56.91 & 51.2  & \cellcolor{green!25}85.7 &
20 & nemeth23     & 72.60 & 67.7  & \cellcolor{green!25}109.2 &
35 & raefsky1     & \cellcolor{green!25}67.28 & 56.4 & 56.8 &
50 & bcsstk15     & 24.05 & \cellcolor{green!25}28.5 & 28.3 \\

6  & TFS b300 c1  & 43.74 & 81.5  & \cellcolor{green!25}91.8 &
21 & nemeth24     & 72.56 & 68.9  & \cellcolor{green!25}108.3 &
36 & psmigr\_3    & 62.76 & 60.8  & \cellcolor{green!25}78.3 &
51 & ex9          & 22.97 & 25.5  & \cellcolor{green!25}25.6 \\

7  & TFS b300     & 43.74 & 81.4  & \cellcolor{green!25}85.3 &
22 & raefsky3     & 34.58 & 47.3  & \cellcolor{green!25}101.5 &
37 & psmigr\_1    & 62.84 & 60.7  & \cellcolor{green!25}78.6 &
52 & Goodwin017   & 22.03 & 24.5  & \cellcolor{green!25}25.0 \\

8  & TRS b678 c1  & 62.56 & 84.9  & \cellcolor{green!25}127.4 &
23 & heart3       & \cellcolor{green!25}90.40 & 87.7  & 88.0 &
38 & psmigr\_2    & 62.64 & 60.5  & \cellcolor{green!25}77.6 &
53 & epb1         & 4.97  & 16.0  & \cellcolor{green!25}16.5 \\

9  & sme3Db       & 36.68 & 105.2 & \cellcolor{green!25}105.5 &
24 & heart2       & \cellcolor{green!25}90.76 & 87.7  & 89.5 &
39 & gyro         & 18.98 & 52.5  & \cellcolor{green!25}89.7 &
54 & dw8192       & \cellcolor{green!25}10.80 & 9.5  & 8.9 \\

10 & smt          & 63.70 & 52.3  & \cellcolor{green!25}124.5 &
25 & raefsky4     & 39.20 & 47.5  & \cellcolor{green!25}97.4 &
40 & nemeth10     & 37.02 & 57.1  & \cellcolor{green!25}62.0 &
55 & dendrimer    & \cellcolor{green!25}24.61 & 18.1 & 17.7 \\

11 & nd3k         & 83.82 & \cellcolor{green!25}128.5 & 128.3 &
26 & msc10848     & 51.36 & 101.7 & \cellcolor{green!25}102.4 &
41 & crystk01     & 41.08 & \cellcolor{green!25}57.4 & 57.0 &
56 & Goodwin013   & \cellcolor{green!25}17.24 & 15.6 & 15.7 \\

12 & Trec14       & 78.10 & 93.5  & \cellcolor{green!25}127.9 &
27 & nemeth21     & 66.01 & 68.3  & \cellcolor{green!25}100.5 &
42 & s2rmq4m1     & 38.30 & \cellcolor{green!25}51.2 & 50.7 &
57 & nasa1824     & \cellcolor{green!25}13.69 & 11.6 & 11.2 \\

13 & heart1       & 96.88 & 110.2 & \cellcolor{green!25}110.8 &
28 & msc23052     & 25.46 & 80.7  & \cellcolor{green!25}88.3 &
43 & bcsstk28     & 36.44 & \cellcolor{green!25}45.6 & 45.3 &
58 & plbuckle     & \cellcolor{green!25}12.25 & 9.4 & 9.4 \\

14 & c8\_mat11    & 90.09 & 111.9 & \cellcolor{green!25}125.5 &
29 & bcsstk36     & 25.33 & 46.1  & \cellcolor{green!25}87.8 &
44 & s3rmt3m3     & 31.69 & 42.6  & \cellcolor{green!25}42.9 &
59 & tomography   & \cellcolor{green!25}13.26 & 9.0 & 9.0 \\

15 & exdata\_1    & 86.42 & 106.5 & \cellcolor{green!25}109.0 &
30 & gyro\_k      & 37.97 & 52.4  & \cellcolor{green!25}89.4 &
45 & t2d\_q9      & 16.13 & \cellcolor{green!25}17.9  & 17.8 &
   &              &       &       &      \\

\hline
\end{tabular}%
}
\caption{Performance comparison in GFLOPs between HiHiSpMV (Denoted HiHi, which does use shuffling) and SUSpMV (NoSh = no shuffle, Sh = shuffle). Green cells indicate the best performance per matrix. The matrices used are from the SuiteSparse Matrix Benchmark set~\cite{suitesparse}. The matrices are sorted by absolute number of nonzeros descending.}
\label{tab:evalThroughputComparison}
\end{table*}

\cref{tab:evalThroughputComparison} shows the computational throughput in GFLOPs for all matrices sorted from highest number of nonzeros to lowest.
Compared to HiHiSpMV, we achieve a geomean speedup of 45\% without shuffling, and 79\% with.
Speedups are more pronounced for matrices with higher number of nonzeros.
For very small matrices, our implementation is in fact slower due to constant factors, such as pipeline depth and arbitration of the Y writer, dominating.

We find a close correlation between the size of the largest CU matrix storage and overall runtime. 
E.g. for \textit{human\_gene1} there are at most 133.8k 256-bit blocks for a single CU with overall runtime at about 137k cycles.

At \MHz{400}, we have a theoretical HBM bandwidth limitation of \GBs{409.6}, while the bandwidth we achieve throughout our benchmarks is \GBs{214.8} (mean), or 52\% and \GBs{387} peak.
When excluding control signals, indices, and inserted zeroes, we are left with an effective bandwidth utilization of \GBs{155.2} for matrix values only.

\subsection{Resource Utilization}

\Cref{tab:resources} shows the resource utilization and peak reported computational throughput for different works using the Alveo U280.
SUSpMV uses significantly more logic and memory resources than HiHiSpMV with the exception of DSPs.
However, since we use 256-bit interfaces on all HBM channels compared to 512-bit for the other implementations, we do not reach the maximum available bandwidth despite achieving higher $f_{max}$.
Storing vector data in a separate DDR memory provides SUSpMV with additional bandwidth and mitigates this drawback.

While this analysis also applies to HiSpMV, it only uses 20 HBM channels for matrix data.
When dividing by the number of HBM channels, both implementations use similar amounts of combined SRAM resources (\texttt{BRAM+URAM*8}).

In 24 of the tests, SUSpMV reaches a higher computational throughput than the reported peak for any previous work on the same platform.

\begin{table}[h]
\centering
\begin{tabularx}{\linewidth}{l|l|l|X|X|l|X}
\hline
Design & LUT & DSP & BRAM & URAM & $f_{max}$ & GFLOPs \\
\hline
Jain et. al.~\cite{leam} & 38\% &  7\% & 23\% & 40\% & 310 & 50.6  \\
Serpens~\cite{serpens}   & 15\% &  8\% & 36\% & 40\% & 270 & 60.55 \\
HiSpMV~\cite{hispmv}     & 56\% & 16\% & 60\% & 33\% & 225 & 50.75 \\
HiHiSpMV~\cite{hihispmv} & 30\% & 27\% & 57\% & 13\% & 225 & 93.69 \\
\textbf{SUSpMV}          & 47\% & 10\% & 82\% & 53\% & 400 & 144.9 \\
\hline
\end{tabularx}
\caption{Resource utilization and $f_{max}$ on Alveo U280. The resource utilization is denoted as a percentage of the total resources available on the whole FPGA.\removed{(So including Board Support Package)}}
\label{tab:resources}
\end{table}

\subsection{CU Scaling}

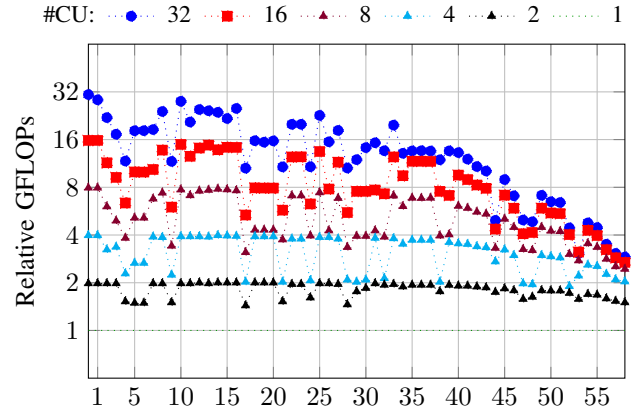
\begin{figure}
\centering
\begin{tikzpicture}
\begin{axis}[
    width=\linewidth,
    height=6cm,
    xlabel={},
    ylabel={Relative GFLOPs},
    grid=major,
    xmin=0, xmax=58,
    ymin=0.5, ymax=64,
    ymode=log, log ticks with fixed point,
    xtick={1,5,10,15,20,25,30,35,40,45,50,55},
    ytick={1, 2, 4, 8, 16, 32},
    legend style={
        at={(0.45,1.02)},
        anchor=south,
        legend columns=7,
        column sep=0.1cm,
        draw=none,
        font=\small,
    }
]
\addlegendimage{empty legend}
\addlegendentry{\#CU:}

\addplot[
    color=blue,
    mark=*,
    dotted
] coordinates {
(0, 30.765957446808507) (1, 28.51063829787234) (2, 22.02127659574468) (3, 17.29787234042553) (4, 11.717391304347826) (5, 18.19148936170213) (6, 18.212765957446805) (7, 18.51063829787234) (8, 24.066666666666666) (9, 11.65217391304348) (10, 27.893617021276594) (11, 20.659574468085104) (12, 24.760869565217394) (13, 24.319148936170212) (14, 23.782608695652176) (15, 21.73913043478261) (16, 25.15217391304348) (17, 10.56521739130435) (18, 15.739130434782611) (19, 15.413043478260873) (20, 15.695652173913045) (21, 10.760869565217392) (22, 19.97826086956522) (23, 19.97826086956522) (24, 10.782608695652176) (25, 22.73913043478261) (26, 15.5) (27, 18.282608695652176) (28, 10.586956521739133) (29, 11.956521739130435) (30, 14.23913043478261) (31, 15.282608695652174) (32, 13.608695652173914) (33, 19.73913043478261) (34, 13.11111111111111) (35, 13.56521739130435) (36, 13.608695652173914) (37, 13.543478260869566) (38, 11.934782608695652) (39, 13.577777777777778) (40, 13.244444444444445) (41, 12.045454545454545) (42, 10.818181818181818) (43, 10.136363636363637) (44, 4.948717948717949) (45, 8.953488372093023) (46, 7.071428571428571) (47, 4.975) (48, 4.85) (49, 7.119047619047619) (50, 6.487804878048782) (51, 6.414634146341464) (52, 4.435897435897436) (53, 3.0606060606060606) (54, 4.76923076923077) (55, 4.45945945945946) (56, 3.5) (57, 3.0625) (58, 2.90625)

};
\addlegendentry{32}

\addplot[
    color=red,
    mark=square*,
    dotted
] coordinates {
(0, 15.808510638297872) (1, 15.808510638297872) (2, 11.425531914893618) (3, 9.23404255319149) (4, 6.369565217391305) (5, 10.0) (6, 9.978723404255318) (7, 10.361702127659575) (8, 13.733333333333333) (9, 6.000000000000001) (10, 14.914893617021274) (11, 12.553191489361701) (12, 14.152173913043478) (13, 14.76595744680851) (14, 13.804347826086957) (15, 14.304347826086957) (16, 14.282608695652176) (17, 5.347826086956522) (18, 7.934782608695653) (19, 7.91304347826087) (20, 7.91304347826087) (21, 5.739130434782608) (22, 12.434782608695654) (23, 12.456521739130435) (24, 6.282608695652174) (25, 13.456521739130435) (26, 7.804347826086957) (27, 11.521739130434783) (28, 5.543478260869565) (29, 7.543478260869566) (30, 7.543478260869566) (31, 7.695652173913044) (32, 7.282608695652175) (33, 12.434782608695654) (34, 9.466666666666667) (35, 11.673913043478262) (36, 11.673913043478262) (37, 11.65217391304348) (38, 7.543478260869566) (39, 7.111111111111111) (40, 9.555555555555555) (41, 8.954545454545453) (42, 8.272727272727272) (43, 7.886363636363637) (44, 4.3589743589743595) (45, 7.1395348837209305) (46, 5.904761904761905) (47, 4.075) (48, 4.15) (49, 5.904761904761905) (50, 5.51219512195122) (51, 5.463414634146342) (52, 4.0256410256410255) (53, 3.1212121212121215) (54, 4.282051282051282) (55, 3.972972972972973) (56, 3.235294117647059) (57, 2.8749999999999996) (58, 2.6874999999999996)
};
\addlegendentry{16}

\addplot[
    color=purple!70!black,
    mark=triangle*,
    dotted
] coordinates {
(0, 7.936170212765957) (1, 7.936170212765957) (2, 6.042553191489361) (3, 4.914893617021277) (4, 3.8260869565217397) (5, 5.127659574468085) (6, 5.127659574468085) (7, 6.76595744680851) (8, 7.377777777777778) (9, 3.4130434782608696) (10, 7.723404255319148) (11, 7.106382978723404) (12, 7.586956521739131) (13, 7.680851063829787) (14, 7.804347826086957) (15, 7.739130434782609) (16, 7.630434782608696) (17, 3.1086956521739135) (18, 4.304347826086957) (19, 4.304347826086957) (20, 4.304347826086957) (21, 3.739130434782609) (22, 7.086956521739132) (23, 7.086956521739132) (24, 3.956521739130435) (25, 7.41304347826087) (26, 4.260869565217392) (27, 6.826086956521739) (28, 3.347826086956522) (29, 3.9347826086956528) (30, 3.9347826086956528) (31, 4.260869565217392) (32, 3.891304347826087) (33, 7.086956521739132) (34, 6.066666666666666) (35, 6.847826086956522) (36, 6.826086956521739) (37, 6.847826086956522) (38, 3.956521739130435) (39, 4.022222222222222) (40, 6.088888888888889) (41, 5.909090909090908) (42, 5.590909090909091) (43, 5.431818181818181) (44, 3.307692307692308) (45, 5.046511627906977) (46, 4.476190476190476) (47, 3.25) (48, 3.2) (49, 4.476190476190476) (50, 4.2439024390243905) (51, 4.219512195121951) (52, 3.025641025641026) (53, 2.757575757575758) (54, 3.5384615384615388) (55, 3.3513513513513513) (56, 2.823529411764706) (57, 2.5312499999999996) (58, 2.4375)

};
\addlegendentry{8}

\addplot[
    color=cyan,
    mark=triangle*,
    dotted
] coordinates {
(0, 3.978723404255319) (1, 3.978723404255319) (2, 3.234042553191489) (3, 3.3617021276595747) (4, 2.2826086956521743) (5, 2.6595744680851063) (6, 2.6595744680851063) (7, 3.893617021276596) (8, 3.8666666666666663) (9, 2.239130434782609) (10, 3.914893617021276) (11, 3.914893617021276) (12, 3.91304347826087) (13, 3.893617021276596) (14, 3.9782608695652177) (15, 3.956521739130435) (16, 3.9347826086956528) (17, 2.0217391304347827) (18, 3.91304347826087) (19, 3.91304347826087) (20, 3.91304347826087) (21, 2.0217391304347827) (22, 3.782608695652174) (23, 3.782608695652174) (24, 2.065217391304348) (25, 3.8695652173913047) (26, 3.8695652173913047) (27, 3.782608695652174) (28, 2.0869565217391304) (29, 2.0217391304347827) (30, 2.0869565217391304) (31, 3.8260869565217397) (32, 2.130434782608696) (33, 3.804347826086957) (34, 3.511111111111111) (35, 3.7173913043478266) (36, 3.7173913043478266) (37, 3.7173913043478266) (38, 2.0217391304347827) (39, 3.5999999999999996) (40, 3.533333333333333) (41, 3.5) (42, 3.3863636363636362) (43, 3.3409090909090904) (44, 2.717948717948718) (45, 3.232558139534884) (46, 2.9761904761904763) (47, 1.975) (48, 1.95) (49, 2.9761904761904763) (50, 2.9268292682926833) (51, 2.878048780487805) (52, 1.8974358974358976) (53, 2.2121212121212124) (54, 2.58974358974359) (55, 2.5405405405405403) (56, 2.264705882352941) (57, 2.09375) (58, 2.03125)
};
\addlegendentry{4}

\addplot[
    color=black,
    mark=triangle*,
    dotted
] coordinates {
(0, 1.9787234042553192) (1, 1.9787234042553192) (2, 1.9787234042553192) (3, 1.9787234042553192) (4, 1.5217391304347827) (5, 1.4893617021276595) (6, 1.4893617021276595) (7, 1.9787234042553192) (8, 1.9777777777777779) (9, 1.5000000000000002) (10, 1.9787234042553192) (11, 1.9787234042553192) (12, 2.0) (13, 1.9787234042553192) (14, 2.0) (15, 2.0) (16, 2.0) (17, 1.434782608695652) (18, 2.0) (19, 2.0) (20, 2.0) (21, 1.5217391304347827) (22, 1.956521739130435) (23, 1.956521739130435) (24, 1.6086956521739133) (25, 1.9782608695652175) (26, 1.9782608695652175) (27, 1.956521739130435) (28, 1.456521739130435) (29, 1.7608695652173914) (30, 1.847826086956522) (31, 1.9782608695652175) (32, 1.9347826086956523) (33, 1.956521739130435) (34, 1.8888888888888888) (35, 1.9347826086956523) (36, 1.9347826086956523) (37, 1.9347826086956523) (38, 1.7608695652173914) (39, 1.9333333333333331) (40, 1.911111111111111) (41, 1.909090909090909) (42, 1.8863636363636365) (43, 1.8636363636363633) (44, 1.7435897435897436) (45, 1.8372093023255816) (46, 1.7857142857142856) (47, 1.575) (48, 1.625) (49, 1.7857142857142856) (50, 1.7804878048780488) (51, 1.7804878048780488) (52, 1.717948717948718) (53, 1.575757575757576) (54, 1.6923076923076923) (55, 1.6756756756756757) (56, 1.5882352941176472) (57, 1.53125) (58, 1.4999999999999998)
};
\addlegendentry{2}

\addplot[
    color=green!60!black,
    dotted
] coordinates {
(0, 1.0) (1, 1.0) (2, 1.0) (3, 1.0) (4, 1.0) (5, 1.0) (6, 1.0) (7, 1.0) (8, 1.0) (9, 1.0) (10, 1.0) (11, 1.0) (12, 1.0) (13, 1.0) (14, 1.0) (15, 1.0) (16, 1.0) (17, 1.0) (18, 1.0) (19, 1.0) (20, 1.0) (21, 1.0) (22, 1.0) (23, 1.0) (24, 1.0) (25, 1.0) (26, 1.0) (27, 1.0) (28, 1.0) (29, 1.0) (30, 1.0) (31, 1.0) (32, 1.0) (33, 1.0) (34, 1.0) (35, 1.0) (36, 1.0) (37, 1.0) (38, 1.0) (39, 1.0) (40, 1.0) (41, 1.0) (42, 1.0) (43, 1.0) (44, 1.0) (45, 1.0) (46, 1.0) (47, 1.0) (48, 1.0) (49, 1.0) (50, 1.0) (51, 1.0) (52, 1.0) (53, 1.0) (54, 1.0) (55, 1.0) (56, 1.0) (57, 1.0) (58, 1.0)
};
\addlegendentry{1}

\end{axis}
\end{tikzpicture}
\caption{CU Scaling comparison. The Y axis shows relative performance to the 1CU version for 2, 4, 8, 16, and 32 CUs respectively}
\label{fig:evalCUScaling}
\end{figure}

To test the scaling of our design with an increasing number of CUs, we perform tests with 1, 2, 4, 8, 16, and 32 CUs.
We perform them by disabling the other CUs by assigning zero Y values to them.
This does leave the constant-time overhead of synchronizing on the Y return channel.
From \cref{fig:evalCUScaling} it is clear to that performance scaling is mostly linear in the number of CUs, with the exception of very small matrices. 

\subsection{Data Format}

The purpose of the 6xFloat format is to take advantage of the denser areas in the matrix, by reducing the bits for indices in order to fit in one more float per block.
We evaluate its effect by disabling 6xFloat blocks and measuring the difference in both the total number of blocks issued and overall runtime.

Results show a 16\% (geomean) reduction in number of blocks and of 12\% (geomean) speedup.
We note that especially larger matrices come close to a 20\% improvement, the maximum possible with this format.

These results indicate that dynamically adapting the index width can significantly increase the efficiency of bandwidth utilization.
While the 6xFloat format works for many entries on the same or consecutive rows, the 5xFloat format only needs to add additional dummy entries when density falls below one entry per 255 rows (about $4*10^{-6}$).
While none of the matrices of our test set have such low-density, many have large empty sections.
When shuffling for a more even distribution is not an option, these empty sections incur a significant overhead in 5xFloat dummy blocks.

While 6xFloat improves computational throughput for dense areas, very sparse areas may require dummy values to bridge the gap.
Over all test matrices, we find that our format requires 0.64\% (geomean) and 6.41\% (worst case) dummy zeroes for any reason, including bank conflicts, but mostly sparse areas.

This indicates that a 4xFloat type could improve performance for some matrices with large empty areas.
Such a type brings its own issues though, as a 4xFloat could skip many tiles at once.
This guarantees desynchronization of CUs and will stall the faster CUs immediately.
To realize the potential of a 4xFloat, X vector reloading must be adapted to such desynchronization.

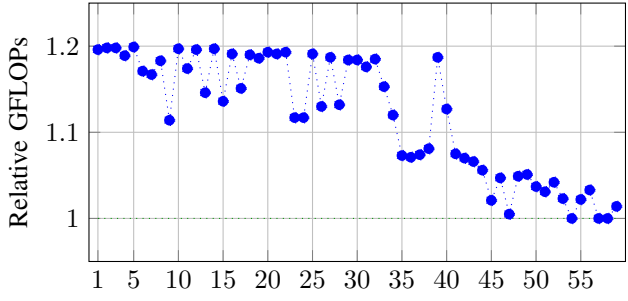
\begin{figure}
\centering
\begin{tikzpicture}
\begin{axis}[
    width=\linewidth,
    height=5cm,
    xlabel={},
    ylabel={Relative GFLOPs},
    legend pos=north east,
    grid=major,
    xmin=0, xmax=60,
    ymin=0.95, ymax=1.25,
    xtick={1,5,10,15,20,25,30,35,40,45,50,55},
    ytick={1.0,1.1,1.2}
]

\addplot[
    color=blue,
    mark=*,
    dotted
] coordinates {
(1, 1.196) (2, 1.198) (3, 1.198) (4, 1.189) (5, 1.199) (6, 1.171) (7, 1.167) (8, 1.183) (9, 1.114) (10, 1.197) (11, 1.174) (12, 1.196) (13, 1.146) (14, 1.197) (15, 1.136) (16, 1.191) (17, 1.151) (18, 1.190) (19, 1.186) (20, 1.193) (21, 1.191) (22, 1.193) (23, 1.117) (24, 1.117) (25, 1.191) (26, 1.130) (27, 1.187) (28, 1.132) (29, 1.184) (30, 1.184) (31, 1.176) (32, 1.185) (33, 1.153) (34, 1.120) (35, 1.073) (36, 1.071) (37, 1.074) (38, 1.081) (39, 1.187) (40, 1.127) (41, 1.075) (42, 1.070) (43, 1.066) (44, 1.056) (45, 1.021) (46, 1.047) (47, 1.005) (48, 1.049) (49, 1.051) (50, 1.037) (51, 1.031) (52, 1.042) (53, 1.023) (54, 1.000) (55, 1.022) (56, 1.033) (57, 1.000) (58, 1.000) (59, 1.014)
};

\addplot[
    color=green!60!black,
    dotted
] coordinates {
(1, 1.0) (2, 1.0) (3, 1.0) (4, 1.0) (5, 1.0) (6, 1.0) (7, 1.0) (8, 1.0) (9, 1.0) (10, 1.0) (11, 1.0) (12, 1.0) (13, 1.0) (14, 1.0) (15, 1.0) (16, 1.0) (17, 1.0) (18, 1.0) (19, 1.0) (20, 1.0) (21, 1.0) (22, 1.0) (23, 1.0) (24, 1.0) (25, 1.0) (26, 1.0) (27, 1.0) (28, 1.0) (29, 1.0) (30, 1.0) (31, 1.0) (32, 1.0) (33, 1.0) (34, 1.0) (35, 1.0) (36, 1.0) (37, 1.0) (38, 1.0) (39, 1.0) (40, 1.0) (41, 1.0) (42, 1.0) (43, 1.0) (44, 1.0) (45, 1.0) (46, 1.0) (47, 1.0) (48, 1.0) (49, 1.0) (50, 1.0) (51, 1.0) (52, 1.0) (53, 1.0) (54, 1.0) (55, 1.0) (56, 1.0) (57, 1.0) (58, 1.0) (59, 1.0)
};

\end{axis}
\end{tikzpicture}
\caption{Ablation study of 6xFloat. The baseline curve is using only 5xFloat blocks, whereas the improved curve is when using both. Observe that performance improves by up to 20\%.}
\label{fig:eval6xFloatAblate}
\end{figure}

\subsection{Accumulator}

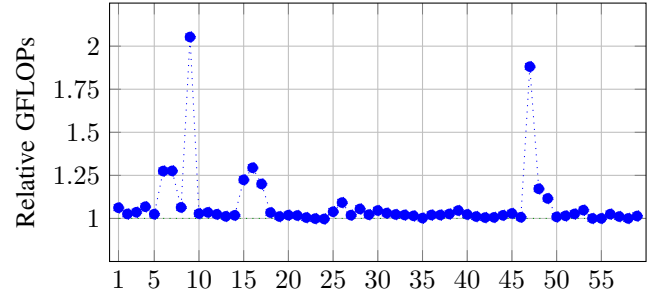
\begin{figure}
\centering
\begin{tikzpicture}
\begin{axis}[
    width=\linewidth,
    height=5cm,
    xlabel={},
    ylabel={Relative GFLOPs},
    legend pos=south east,
    grid=major,
    xmin=0, xmax=60,
    ymin=0.75, ymax=2.25,
    xtick={1,5,10,15,20,25,30,35,40,45,50,55},
    ytick={1.0,1.25,1.5, 1.75,2.0}
]

\addplot[
    color=blue,
    mark=*,
    dotted
] coordinates {
(1, 1.062) (2, 1.026) (3, 1.036) (4, 1.068) (5, 1.025) (6, 1.275) (7, 1.276) (8, 1.064) (9, 2.053) (10, 1.028) (11, 1.035) (12, 1.024) (13, 1.012) (14, 1.018) (15, 1.224) (16, 1.294) (17, 1.200) (18, 1.034) (19, 1.011) (20, 1.019) (21, 1.017) (22, 1.005) (23, 0.999) (24, 0.997) (25, 1.039) (26, 1.092) (27, 1.019) (28, 1.055) (29, 1.022) (30, 1.046) (31, 1.031) (32, 1.023) (33, 1.020) (34, 1.015) (35, 1.002) (36, 1.020) (37, 1.020) (38, 1.026) (39, 1.046) (40, 1.023) (41, 1.011) (42, 1.005) (43, 1.006) (44, 1.018) (45, 1.029) (46, 1.007) (47, 1.881) (48, 1.172) (49, 1.116) (50, 1.009) (51, 1.015) (52, 1.026) (53, 1.048) (54, 1.000) (55, 1.000) (56, 1.025) (57, 1.011) (58, 1.000) (59, 1.014)
};

\addplot[
    color=green!60!black,
    dotted
] coordinates {
(1, 1.0) (2, 1.0) (3, 1.0) (4, 1.0) (5, 1.0) (6, 1.0) (7, 1.0) (8, 1.0) (9, 1.0) (10, 1.0) (11, 1.0) (12, 1.0) (13, 1.0) (14, 1.0) (15, 1.0) (16, 1.0) (17, 1.0) (18, 1.0) (19, 1.0) (20, 1.0) (21, 1.0) (22, 1.0) (23, 1.0) (24, 1.0) (25, 1.0) (26, 1.0) (27, 1.0) (28, 1.0) (29, 1.0) (30, 1.0) (31, 1.0) (32, 1.0) (33, 1.0) (34, 1.0) (35, 1.0) (36, 1.0) (37, 1.0) (38, 1.0) (39, 1.0) (40, 1.0) (41, 1.0) (42, 1.0) (43, 1.0) (44, 1.0) (45, 1.0) (46, 1.0) (47, 1.0) (48, 1.0) (49, 1.0) (50, 1.0) (51, 1.0) (52, 1.0) (53, 1.0) (54, 1.0) (55, 1.0) (56, 1.0) (57, 1.0) (58, 1.0) (59, 1.0) 
};

\end{axis}
\end{tikzpicture}
\caption{Ablation study of accumulating multiple rows per block. The baseline is the performance if at most one row can be accumulated per block, and the improved version allows up to 6 rows to be accumulated per block. \removed{Matrices "sme3Db" and "ted\_B" stand out, due to a large number of 1-element rows.} }
\label{fig:evalMultiRowAblate}
\end{figure}

Compared to previous designs, our accumulator can accumulate values of different rows separately within the same cycle.
We purposefully alter the matrix storage format generation in software to only insert values of the same row into a block and move other values to the next.
We thus measure the impact of our novel accumulator design as shown in \cref{fig:evalMultiRowAblate}.
This leads to an overall throughput improvement of 7.9\% (geomean) with a best case of 102\%.

The two most extreme outliers are \textit{sme3Db} and \textit{ted\_B}.
A closer look shows this to be due to unequal distribution of matrix data blocks.
While each CU receives roughly the same amount of matrix entries, those entries are spread across different amounts of rows.
In the case of \textit{ted\_B}, the first 4k rows go to CU0, while the remaining CUs handle about 200 rows each.
Nonetheless, enabling the full accumulator still reduced the matrix storage size from 30k to 24k blocks.
Over all matrices, we see a storage size reduction of 7.6\% (geomean).

\section{Conclusion}

In this work, we presented SUSpMV, a novel SpMV accelerator implemented in the HDL SUS and released as open-source software \ifblinded\else \cite{suspmv-github-do-not-use} \fi.
We compared SUSpMV against HiHiSpMV on the same Alveo U280 platform and achieve 79\% higher computational throughput.

Our ablation study showed that our main contributions both significantly improve overall performance.
First, the dynamic matrix storage format increases data delivery bandwidth to the accelerator by up to 20\% in dense matrix regions and achieves this on a large variety of matrices.
Second, the novel multi-row accumulator further reduces matrix storage size by 7.6\%, increasing overall effective bandwidth.

Our evaluation further demonstrated good scalability with increasing numbers of CUs for larger matrices.
These results highlight the potential of future implementations on Versal HBM FPGAs, which provide up to 64 HBM channels.
In addition, their dedicated fp32 DSP resources would significantly reduce the LUT and FF utilization of our design.

Finally, our study identified further possible improvements to the dynamic matrix storage encoding and more efficient streaming of the X vector could unlock additional performance gains in future work.

\section*{Acknowledgment}

\ifblinded
\else
This research was funded by the German Research Foundation (DFG) in project \#419942270 neoDBMS.
The authors gratefully acknowledge the funding of this project by computing time provided by the Paderborn Center for Parallel Computing (PC2).
The authors would like to thank AMD for supporting their work by donations of hard- and software.
\fi

\bibliographystyle{IEEEtran}
\IEEEtriggeratref{14}
\bibliography{references}
\end{document}